\documentclass[twocolumn,english]{revtex4-2}
\usepackage[T1]{fontenc}
\usepackage[utf8]{inputenc}
\usepackage{amsmath}
\usepackage{graphicx}

\makeatletter
\usepackage[utf8]{inputenc}
\usepackage[T1]{fontenc}

\DeclareUnicodeCharacter{0229}{\c{e}} 

\makeatother

\usepackage{babel}
\begin{document}
\title{Diffusion of rod-like particles in complex fluids}
\author{Władysław Sokołowski}
\author{Huma Jamil}
\author{Karol Makuch}
\email{Electronic Address: E-mail address: kmakuch@ichf.edu.pl}

\affiliation{Institute of Physical Chemistry, Polish Academy of Sciences}
\address{ul. Kasprzaka 44/52, 01-224 Warszawa, Poland}
\begin{abstract}
Diffusion of particles in complex fluids and gels is difficult to
describe and often lies beyond the scope of the classical Stokes--Einstein
relation. One of the main lines of research over the past few decades
has sought to relate diffusivity to a fundamental dissipative property
of the fluid: the wave-vector-dependent shear viscosity function.
Here, we use linear response theory to extend this viscosity function
framework to rod-like particles. Using a dimer (two-bead particle)
as a minimal rod-like probe, we derive explicit expressions for its
diffusion coefficients parallel and perpendicular to its axis in terms
of the viscosity function. We show that this description captures
the full range of behaviors, from nearly isotropic diffusion of the
rod-like probe to highly anisotropic, reptation-like motion. The method
is based on a microscopic statistical-mechanical treatment of the
Smoluchowski dynamics, yet leads to simple final formulas, providing
a practical tool for interpreting diffusion experiments on rod-like
tracers in complex fluids. We also clarify the limitations of this
approach, emphasizing that the present formulation is primarily suited
to complex liquids like polymer solutions and only indirectly applicable
to gels.
\end{abstract}
\maketitle

\section{Introduction}

Many macromolecules in nature possess a rod-like shape. This class
includes actin filaments, microtubules, DNA fragments, certain viruses,
cellulose fibers, and various synthetic nanorods \citep{CelluloseNano,MotorProteins,NucleicAcids}.
In most of these cases, such rod-like particles exist in a liquid
medium - either within the crowded interior of biological cells, which
host thousands of molecular species, or in engineered environments
such as liquid crystals used in industry \citep{Masao_Doi_Soft_Matter_Physics}.

Understanding macromolecular motion is thus of great practical and
fundamental importance, particularly when the system is close to equilibrium.
Even in the absence of external forces, a macromolecule is constantly
jostled by surrounding atoms and molecules, causing its velocity to
change direction repeatedly. At long times, this random motion becomes
diffusive, with the mean-square displacement obeying $\left\langle \left[{\bf R}\left(t\right)-{\bf R}\left(0\right)\right]^{2}\right\rangle =6Dt$
which defines the diffusion coefficient $D$. The classical work of
Sutherland, Einstein, and Smoluchowski established how this thermal
motion is connected to dissipation. This result, known as the Einstein
relation, $D=k_{B}T\mu$, relates the diffusion coefficient to the
mobility $\mu$ measured from velocity response, ${\bf V}=\mu{\bf F}$,
to a small applied force ${\bf F}$. In the Einstein formula, $k_{B}T$
is the Boltzmann constant and the absolute temperature. Because this
relation follows from general principles of linear response theory,
any systematic deviation in experiment or simulation indicates either
a breakdown of equilibrium assumptions or an inconsistency in the
methodology.

Einstein's relation thus provides a practical route for understanding
diffusivity by studying the mobility of a probe subjected to a small
force - the approach adopted in this work. Mobility quantifies the
rate at which the work done by the applied force, ${\bf F}\cdot{\bf V}={\bf F}\cdot\mu{\bf F}$,
is dissipated in the surrounding fluid. A moving probe stores essentially
no energy; instead, the energy input is continuously dissipated through
viscous shear in the medium. Consequently, it is not surprising that
the mobility of a spherical particle of radius $a$ in a Newtonian
fluid, $\mu=1/6\pi\eta_{0}a,$ is determined by the fluid’s shear
viscosity $\eta_{0}$.

The dissipative (viscous) properties of a fluid can be probed by applying
a sinusoidal volumetric force density acting on all fluid particles,
${\bf f}\left({\bf r}\right)=f_{0}{\bf e}_{x}\exp\left(-ik{\bf e}_{z}\cdot{\bf r}\right)$.
According to linear response theory, such forcing generates a velocity
field of the same form, ${\bf v}\left({\bf r}\right)=v_{0}{\bf e}_{x}\exp\left(-ik{\bf e}_{z}\cdot{\bf r}\right)$,
with amplitude $v_{0}=f_{0}/k^{2}\eta\left(k\right)$. The function
$\eta\left(k\right)$, known as the wave-vector--dependent shear
viscosity \citep{cite-key} is a fundamental property of any fluid
\citep{Denis_J_Evans_Gary_Morriss_Statistical_Mechanics_of_Nonequilibrium_liquids}.
This relation provides a general route for determining $\eta\left(k\right)$
in atomic, molecular, and complex fluids, including gels. For small
wave-vectors $\eta\left(k\right)$ reduces to the macroscopic shear
viscosity, $\eta_{\text{macro}}=\lim_{k\to0}\eta\left(k\right)$.
For simple molecular liquids, simulations show that $\eta\left(k\right)$
decreases with increasing $k$ and approaches zero at wavelengths
corresponding to only a few angstroms \citep{balucani1987transverse}.
In contrast, the viscosity function of complex fluids has been explored
far less \citep{cite-key,BEENAKKER198448}. In the context of Smoluchowski
dynamics - a coarse-grained description appropriate for colloidal
suspensions and other macromolecular systems - $\eta\left(k\right)$
interpolates between the macroscopic viscosity at small wave-vectors,
$\eta_{\text{macro}}=\lim_{k\to0}\eta\left(k\right)$, and the solvent
viscosity at large wave-vectors, $\eta_{0}=\lim_{k\to\infty}\eta\left(k\right)$.

Linear response theory shows that the shear viscosity function also
governs the velocity field generated by a localized perturbation,
not only by a sinusoidal driving force. The average velocity field,
$\left\langle {\bf v}\left(\mathbf{R}\right)\right\rangle $, of an
incompressible, homogeneous, and isotropic fluid subjected to a small
point force ${\bf f}$ is given by \citep{diag_ostateczna_wersja}
\begin{equation}
\left\langle {\bf v}\left(\mathbf{R}\right)\right\rangle =G_{\text{eff}}\left({\bf R}\right){\bf f},\label{eq:Velocity=000020field}
\end{equation}
where $G_{\text{eff}}\left({\bf R}\right)$ is called the effective
Green function, and has the general Fourier-space form
\begin{equation}
\widehat{G}_{\mathrm{eff}}\left(\mathbf{k}\right)=\frac{1}{k^{2}\eta\left(k\right)}\left(\mathbf{I}-\hat{\mathbf{k}}\hat{\mathbf{k}}\right),\label{eq:=000020Effective=000020Green=000020function}
\end{equation}
explicitly involving the wave-vector--dependent viscosity $\eta\left(k\right)$.
In practice, most numerical studies determine $\eta\left(k\right)$
either from equilibrium autocorrelation functions \citep{gaskell1987wavevector,hansen2007parameterization}
or by applying a sinusoidal force \citep{glavatskiy2015nonlocal}.
The use of a point-force field, as in the expression above, to extract
the viscosity function is rare. Experimental determinations of the
viscosity function remain scarce. A recent approach proposes extracting
$\eta\left(k\right)$ from measurements of the diffusion of probe
particles with different hydrodynamic radii \citep{C9SM01119F}.

We argued above that the particle mobility is linked to the viscosity
function $\eta\left(k\right).$ Determining the exact, general relation
between them is, however, a difficult and still unsolved problem.
This connection has been explored for spherical probes in two principal
contexts: to understand deviations from the Stokes-Einstein relation
in molecular liquids \citep{keyes1973bilinear,keyes1975self,kim2005breakdown},
and in such complex fluids as polymer melts \citep{yamamoto2011theory}
and colloidal suspensions in the limit where hydrodynamic effects
are dominant \citep{BEENAKKER198448}. Within Smoluchowski dynamics,
the mobility can be related to $\eta\left(k\right),$ but direct interactions
between the probe and surrounding macromolecules also contribute \citep{C9SM01119F,everts2025brownian}.
A complete statistical-physics treatment remains challenging and requires
further development \citep{diag_ostateczna_wersja,everts2025brownian}.
Existing analyses nevertheless provide simple phenomenological approximations
that capture the orders-of-magnitude variation of diffusivity across
different probe sizes, reflecting the hierarchy of length scales present
in complex fluids. Comparable theoretical understanding is largely
absent for nonspherical probes such as rod-like particles.

The diffusion of rod-like particles displays richer behavior than
that of spheres. In simple Newtonian fluids, their mobility is anisotropic:
motion along the rod’s axis is easier than motion perpendicular to
it. The mobility ratio $\mu_{\parallel}/\mu_{\perp}$ exceeds unity
and depends logarithmically on the aspect ratio $p=L/d$ (length to
diameter). In the large-$p$ limit, one finds $\mu_{\parallel}/\mu_{\perp}=1+0.09/\log p$
\citep{bitter2017interfacial}. For very slender rods, the difference
between longitudinal and transverse mobility becomes small, and the
center-of-mass diffusion becomes isotropic. In contrast, rod diffusion
in dense polymer melts exhibits the opposite trend: $\mu_{\parallel}/\mu_{\perp}$
increases strongly with $p$ indicating a pronounced suppression of
transverse motion \citep{zhang2025cross}. In this regime, rods predominantly
translate along their long axis while lateral displacements are strongly
hindered - a behavior reminiscent of reptation dynamics originally
introduced for motion between immobile obstacles \citep{de1971reptation}.

In this paper, we ask whether the viscosity function $\eta\left(k\right)$
can also capture the rich dynamical behavior of rod-like particles
in complex fluids.

We compute the mobility of a dimer - the simplest model of a rod-like
particle - by assuming that the dominant hydrodynamic contribution
arises from the coupling of its two beads through the effective Green
function $G_{\text{eff}}\left({\bf R}\right)$. This approach follows
recent statistical-physics developments for complex fluids \citep{C9SM01119F,diag_ostateczna_wersja}.
Under this assumption, we obtain a simple approximation reminiscent
of Smoluchowski’s original treatment of two interacting spheres, yielding
explicit expressions for the parallel and perpendicular mobilities
of the dimer in terms of the viscosity function and the hydrodynamic
radii of its constituent beads.

This framework allows us to evaluate how a wave-vector-dependent viscosity
influences the anisotropic motion of a rod-like probe. Remarkably,
different forms of $\eta\left(k\right)$ naturally reproduce both
spherical-like behavior in simple fluids and reptation-like dynamics
in crowded environments.

A key strength of this approach is that it relies solely on the viscosity
function - a fundamental property of any fluid - without invoking
system-specific microscopic details \citep{quesada2021solute} or
phenomenological fitting parameters \citep{amsden1998solute}. In
contrast to numerical simulations that proceed directly from a model
system to its diffusivity \citep{rokhforouz2025brownian}, our method
highlights the role of $\eta\left(k\right)$ as the underlying physical
quantity controlling mobility.

\section{Approximation}

In many experiments, rod-like particles are tracked by monitoring
their motion along a fixed orientation \citep{han2006brownian}. The
parallel and perpendicular components of mobility then provide valuable
insight into the surrounding medium. To mimic this situation, we consider
a rod-like particle that maintains its orientation and model it as
an oriented dumbbell. The velocity response of such a dumbbell necessarily
takes the form
\begin{equation}
{\bf V}=\mathbf{M}\left({\bf R}\right){\bf F}_{\text{total}},\label{eq:dumbbell=000020mobility=000020matrix}
\end{equation}
where ${\bf R}$ denotes the vector connecting the centers of the
two beads and ${\bf F}_{\text{total}}$ is the total applied force.
Because the surrounding complex fluid is statistically isotropic,
the mobility matrix must be isotropic as well and can therefore be
decomposed as
\begin{equation}
\mathbf{M}\left({\bf R}\right)=\mu_{\parallel}\left(R\right)\,\hat{\mathbf{R}}\hat{\mathbf{R}}+\mu_{\perp}\left(R\right)\left(\mathbf{I}-\hat{\mathbf{R}}\hat{\mathbf{R}}\right),\label{eq:def=000020of=000020mobilities}
\end{equation}
with $\mu_{\parallel}$ and $\mu_{\perp}$ denoting the parallel and
perpendicular mobility coefficients.

For sufficiently large bead separations, hydrodynamic coupling between
the beads becomes negligible. In this limit, the dumbbell behaves
as two independent particles constrained to move together, and its
center-of-mass mobility is simply one-half of the single-bead mobility,
$\mu_{\parallel}\left(R\right)=\mu_{\perp}\left(R\right)=\mu_{\text{single}}/2$.

To analyze $\mu_{\parallel}$ and $\mu_{\perp}$ in more detail, we
work within Smoluchowski dynamics, which provides a coarse-grained
description of macromolecules in solution \citep{dhont2003introduction}.
Unlike molecular dynamics, Smoluchowski dynamics averages over the
solvent degrees of freedom, leaving only the positional degrees of
freedom of the macromolecules. These undergo Brownian motion under
both direct interactions and hydrodynamic interactions mediated by
the solvent.

In this paper, we analyze the parallel and perpendicular mobilities
of a dumbbell through the lens of the effective two-particle mobility
matrix. This matrix arises naturally when considering the motion of
macromolecules subjected to a small external force. A force ${\bf F}_{1}$
applied to particles at position ${\bf R}_{1}$, sets the entire complex
fluid into motion, including nearby macromolecules. For sufficiently
small forces, the average velocity response of particles at a second
position, ${\bf R}_{2}$, is linear,
\[
{\bf V}_{2}=\mu_{21}^{\text{eff}}\left({\bf R}_{2}-{\bf R}_{1}\right){\bf F}_{1},
\]
where $\mu_{21}^{\text{eff}}\left({\bf R}\right)$ is the effective
pair mobility. Likewise, the particles on which the force acts respond
linearly, 
\[
{\bf V}_{1}=\mu_{\text{self}}^{\text{eff}}{\bf F}_{1},
\]
with $\mu_{\text{self}}^{\text{eff}}$ the effective self-mobility
matrix \citep{diag_ostateczna_wersja}.

The effective mobility matrices described above have been analyzed
in various contexts within Smoluchowski dynamics. However, deriving
fully microscopic expressions for these quantities remains challenging
\citep{diag_ostateczna_wersja,everts2025brownian}. What is known
is that the effective pair mobility of two particles has the following
Fourier-space structure, $\hat{\mu}_{12}^{\text{eff}}\left({\bf k}\right)=\hat{\mu}_{12}^{\text{irr}}\left({\bf k}\right)+\hat{\mu}_{<}^{\text{irr}}\left({\bf k}\right)G_{\text{eff}}\left({\bf k}\right)\hat{\mu}_{>}^{\text{irr}}\left({\bf k}\right)$
\citep{diag_ostateczna_wersja}. This relation provides a clear example
of how a macroscopic transport quantity - here, the effective pair
mobility $\hat{\mu}_{12}^{\text{eff}}\left({\bf k}\right)$ - naturally
incorporates the viscosity function through $G_{\text{eff}}$. We
do not examine the details of the matrices $\hat{\mu}_{12}^{\text{irr}}\left({\bf k}\right)$,
$\hat{\mu}_{<}^{\text{irr}}\left({\bf k}\right)$ and $\hat{\mu}_{>}^{\text{irr}}\left({\bf k}\right)$,
it is sufficient to note that the “irr” terms are expected to be short-ranged,
decaying faster than $1/R^{3}$ for large $R$ in real space for Stokes
flow \citep{Cichocki2002three}, and that $\hat{\mu}_{>,<}^{\text{irr}}\left({\bf k}\right)$
reduce to the identity matrix as ${\bf k}\to0$ \citep{everts2025brownian}.
Under these conditions, the long-distance behavior of the effective
pair mobility simplifies to

\begin{equation}
\mu_{12}^{\text{eff}}\left({\bf R}\right)\approx G_{\text{eff}}\left(\mathbf{R}\right),\label{eq:far=000020field=000020app}
\end{equation}
where ${\bf G}_{\text{eff}}\left({\bf R}\right)$ is the effective
Green tensor. This mirrors Smoluchowski’s result for two sedimenting
spheres, where hydrodynamic interactions at large separations were
captured by the bare Oseen tensor, $\mu_{12}^{\text{eff}}\left({\bf R}\right)\approx G_{0}\left(\mathbf{R}\right)$
\citep{Smoluchowski1927}. 

We use the above dominant contribution to describe how the beads in
an oriented dumbbell influence each other’s motion in the mobility
problem. A force ${\bf F}_{2}$ acting on bead 2 generates a hydrodynamic
flow field $G_{\text{eff}}\left(\mathbf{R}\right){\bf F}_{2}$ at
the position of bead 1. Applying the same force to bead 1, ${\bf F}_{1}={\bf F}_{2}={\bf F}_{\text{total}}/2$,
produces an analogous contribution. Combining these effects yields
the following approximation for the dumbbell mobility matrix:
\begin{equation}
\mathbf{M}\left({\bf R}\right)\approx\frac{1}{2}\mu_{\text{single}}{\bf 1}+\frac{1}{2}G_{\text{eff}}\left(\mathbf{R}\right).\label{eq:dominant=000020contribution=000020app}
\end{equation}
This expression constitutes the main approximation scheme introduced
in this work. It requires only two ingredients: the single-bead mobility
matrix and the viscosity function $\eta\left(k\right)$ , which enters
through the effective Green function $G_{\text{eff}}\left({\bf k}\right)$
defined in Eq. (\ref{eq:=000020Effective=000020Green=000020function}).

The limitations of the approximation in Eq. (\ref{eq:dominant=000020contribution=000020app})
ultimately reduce to determining under which conditions the exact
fixed-orientation dumbbell mobility indeed simplifies to this form.
This question can be addressed directly in numerical simulations,
since the approximation depends only on the single-bead mobility and
the viscosity function. A complementary route is to analyze the problem
using rigorous statistical-physics methods. Guided by earlier work
\citep{diag_ostateczna_wersja,Makuch2015generalization,C9SM01119F,everts2025brownian},
we expect the exact mobility of a dumbbell to have the structure,
${\bf M}\left({\bf R}\right)=\mu_{11}^{\text{dl}}\left({\bf R}\right)+\mu_{12}^{\text{dl}}\left({\bf R}\right)$
with $\hat{\mu}_{12}^{\text{dl}}\left({\bf k}\right)=\hat{\mu}_{12}^{\text{dl,irr}}\left({\bf k}\right)+\hat{\mu}_{<}^{\text{dl,irr}}\left({\bf k}\right)G_{\text{eff}}\left({\bf k}\right)\hat{\mu}_{>}^{\text{dl,irr}}\left({\bf k}\right)$
with the viscosity function entering via $G_{\text{eff}}$ and the
properties analogous to $\mu^{\text{irr}}$ matrices discussed above
for effective pair mobility. Current understanding \citep{everts2025brownian}
indicates that direct interactions between the dumbbell and surrounding
macromolecules of a complex fluid can significantly influence the
range of the $\mu^{\text{dl,irr}}\left({\bf R}\right)$.

Nevertheless, when the beads neither adhere to nearby macromolecules
nor strongly perturb their structure through long-range repulsion,
we anticipate that Eq. (\ref{eq:dominant=000020contribution=000020app})
provides a reliable approximation for the dumbbell mobility. Such
conditions exclude situations involving caging or trapping of probes
in gels.

\section{The wave-vector-dependent viscosity, $\eta\left(k\right)$}

One of the two essential components of the approximation given by
Eq. (\ref{eq:dominant=000020contribution=000020app}) is the effective
Green function from Eq. (\ref{eq:=000020Effective=000020Green=000020function}).
Its inverse Fourier transform to position space yields,
\begin{equation}
G_{\text{eff}}\left(\mathbf{R}\right)=\phi\left(R\right)\,\mathbf{I}+\psi\left(R\right)\left(\mathbf{I}-3\hat{\mathbf{R}}\hat{\mathbf{R}}\right),\label{eq:=000020Scalar=000020EG}
\end{equation}
with two scalar functions $\phi\left(R\right)$ and $\psi\left(R\right)$,
\begin{equation}
\phi\left(R\right)=\frac{1}{3\pi{}^{2}}\int_{0}^{\infty}\frac{j_{0}\left(kR\right)}{\eta\left(k\right)}dk,\label{eq:=000020phi=000020with=000020eta(k)}
\end{equation}
\begin{align}
\psi\left(R\right) & =-\frac{1}{6\pi{}^{2}}\int_{0}^{\infty}\frac{j_{2}\left(kR\right)dk}{\eta\left(k\right)},\label{eq:=000020psi=000020with=000020eta(k)}
\end{align}
where $j_{0}\left(kR\right)=\text{sin}\left(kR\right)/kR$ and $j_{2}\left(kR\right)=\left(\frac{3}{\left(kR\right)^{3}}-\frac{1}{kR}\right)\text{sin}\left(kR\right)-\frac{3}{\left(kR\right)^{2}}\text{cos}\left(kR\right)$
are spherical Bessel functions. From incompressibility, together with
the conditions $\lim_{R\to0}\phi\left(R\right)R^{3}=\lim_{R\to0}\psi\left(R\right)R^{3}=0$,
one obtains the following relation between these scalar functions,
\begin{equation}
\psi\left(R\right)=\frac{1}{2}\phi\left(R\right)-\frac{3}{2R^{3}}\int_{0}^{R}s^{2}\phi\left(s\right)ds.\label{eq:psi(R)=000020general}
\end{equation}

\begin{figure}
\includegraphics[width=8.5cm]{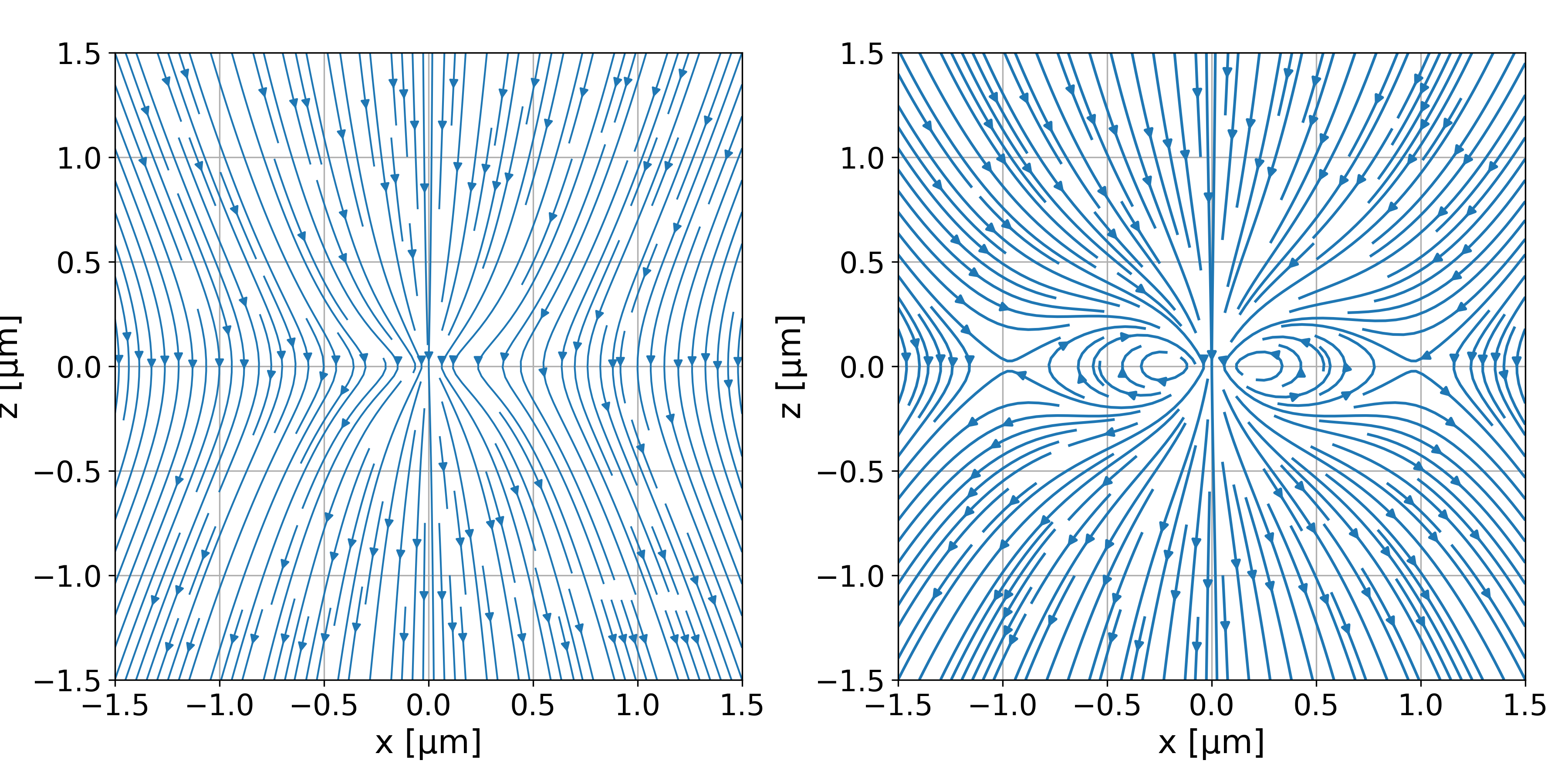}

\caption{ Streamlines of the velocity fields from Eq. (\ref{eq:Velocity=000020field})
generated by a point force $\mathbf{F}=-10^{-12}\hat{\mathbf{z}}\,\text{N}$
with the viscosity function given by Eq. (\ref{eq:=000020rational=000020eta(k)})
for: (a) length $\lambda=10^{-7}\text{m}$, viscosity ratio $\eta_{\text{macro}}/\eta_{0}=5$;
(b) length $\lambda=10^{-7}\text{m}$, viscosity ratio $\eta_{\text{macro}}/\eta_{0}=48$.}\label{fig:Velocity=000020fields}
\end{figure}

A prototypical example of a complex fluid is a suspension of spherical
particles, in which the viscosity function equals the macroscopic
viscosity at $k=0$ and decreases with $k$ approaching the solvent
viscosity $\eta_{0}$ \citep{BEENAKKER198448}. To model this behavior,
we introduce the phenomenological form
\begin{equation}
\eta\left(k\right)=\eta_{\text{macro}}\frac{1+\left(\lambda k\right)^{2}}{1+\frac{\eta_{\text{macro}}}{\eta_{0}}\left(\lambda k\right)^{2}},\label{eq:=000020rational=000020eta(k)}
\end{equation}
which contains the parameters $\eta_{\text{macro}}$, $\eta_{0}$,
and $\lambda$, interpreted as the macroscopic viscosity, the solvent
viscosity, and a length scale governing the crossover from $\eta_{\text{macro}}$
to $\eta_{0}$. Substituting this expression into Eqs. (\ref{eq:=000020phi=000020with=000020eta(k)})
and (\ref{eq:=000020psi=000020with=000020eta(k)}) yields the scalar
functions in the effective Green function:
\begin{equation}
\phi\left(R\right)=\frac{1}{6\pi\eta_{\text{macro}}R}\left[1+\left(\frac{\eta_{\text{macro}}}{\eta_{0}}-1\right)\exp\left(-\frac{R}{\lambda}\right)\right],\label{eq:phen=000020phi}
\end{equation}
\begin{multline}
\psi\left(R\right)=-\frac{1}{12\pi\eta_{\text{macro}}}\Bigg[\left(\frac{\eta_{\text{macro}}}{\eta_{0}}-1\right)\frac{3\lambda^{2}}{R^{3}}+\\
+\frac{1}{2R}-\left(\frac{\eta_{\text{macro}}}{\eta_{0}}-1\right)\exp\left(-\frac{R}{\lambda}\right)\left(\frac{1}{R}+\frac{3\lambda}{R^{2}}+\frac{3\lambda^{2}}{R^{3}}\right)\Bigg].\label{eq:phen=000020psi}
\end{multline}
In the special case where the macroscopic shear viscosity equals the
solvent viscosity, $\eta_{\text{macro}}=\eta_{0}$, the effective
Green function reduces to the Oseen tensor $G_{0}\left(\mathbf{R}\right)=\left(\mathbf{I}+\hat{\mathbf{R}}\hat{\mathbf{R}}\right)/8\pi\eta_{0}R$
\citep{lisicki2013approacheshydrodynamicgreensfunctions}. Since ${\bf f}\cdot G_{0}\left(\mathbf{R}\right)\cdot{\bf {\bf f}}>0$
for all ${\bf R}$, a downward force induces a downward motion of
the fluid everywhere (although the velocity need not be parallel to
the force). We observe qualitatively similar behavior in a complex
fluid with a modest viscosity contrast, for example, $\eta_{\text{macro}}/\eta_{0}=5$
and $\lambda=10^{-7}\text{m}$, as shown in Fig. \ref{fig:Velocity=000020fields}(a).
However, increasing the viscosity ratio $\eta_{\text{macro}}/\eta_{0}$
leads to regions where the local flow reverses direction relative
to the applied force, as illustrated in \ref{fig:Velocity=000020fields}(b)
for $\eta_{\text{macro}}/\eta_{0}=48$ and $\lambda=10^{-7}\text{m}$.
The corresponding velocity field exhibits vortex-like structures around
these regions of reversed flow, reminiscent of similar “vortices”
reported in porous media \citep{vilfan2025stokesdragspherethreedimensional}.

\section{Application}

Using Eq. (\ref{eq:=000020Scalar=000020EG}) in the approximation
(\ref{eq:dominant=000020contribution=000020app}) and identifying
the mobility coefficient through the definition (\ref{eq:def=000020of=000020mobilities})
we obtain
\begin{equation}
\mu_{\parallel}=\frac{1}{2}\left[\mu_{\text{single}}\left(a\right)+\phi\left(R\right)-2\psi\left(R\right)\right],\label{eq:mu=000020parallel}
\end{equation}
\begin{equation}
\mu_{\perp}=\frac{1}{2}\left[\mu_{\text{single}}\left(a\right)+\phi\left(R\right)+\psi\left(R\right)\right].\label{eq:mu=000020perp}
\end{equation}
These expressions constitute a simple methodology for determining
the dumbbell mobilities directly from the single-bead mobility $\mu_{\text{single}}$
and the viscosity function $\eta\left(k\right)$ (equivalently, from
the scalar functions $\phi\left(R\right)$ and $\psi\left(R\right)$).
The difference between the above mobility coefficients is,
\begin{equation}
\mu_{\parallel}-\mu_{\perp}=-\frac{3}{2}\psi\left(R\right).\label{eq:ratio=000020difference}
\end{equation}

Although the methodology outlined above is complete, a determination
of the viscosity function is often cumbersome. To proceed further,
we use a recently proposed phenomenological relation between the self-mobility
of a bead and the viscosity function \citep{C9SM01119F},

\begin{equation}
\mu_{\text{single}}\left(a\right)\approx\frac{1}{3\pi^{2}}\int_{0}^{\infty}\mathrm{d}k\,\frac{j_{0}(ka)}{\eta(k)}.\label{eq:KarolWsOporu}
\end{equation}
This expression leads to simple formulas for the effective Green function.
In particular, inserting Eq. (\ref{eq:=000020phi=000020with=000020eta(k)})
into the above integral yields the positional component
\begin{equation}
\phi\left(R\right)\approx\mu_{\text{single}}\left(R\right),\label{eq:phi=000020within=000020Soft=000020Matter}
\end{equation}
so that $\phi\left(R\right)$ is directly expressed through the self-mobility
of a spherical particle of radius $R$. The second component $\psi\left(R\right)$
then follows from Eq. (\ref{eq:psi(R)=000020general}). 

Applying this phenomenological scheme to the viscosity function in
Eq. (\ref{eq:=000020rational=000020eta(k)}) produces Eqs. (\ref{eq:phen=000020phi})
and (\ref{eq:phen=000020psi}) for $\phi\left(R\right)$ and $\psi\left(R\right)$.
With these ingredients, the “dominant contribution approximation’’
defined by Eq. (\ref{eq:dominant=000020contribution=000020app}) directly
yields the parallel and perpendicular mobilities of the dumbbell.

Equation (\ref{eq:phen=000020phi}) contains an exponential term,
and $\ensuremath{\phi\left(R\right)}$ decreases monotonically with
increasing $R$. A monotonicity analysis of Eq. (\ref{eq:phen=000020psi})
shows that $\psi\left(R\right)$ increases monotonically toward zero
from below; hence, $\psi\left(R\right)<0$ for all $R>0$. Using Eq.
(\ref{eq:ratio=000020difference}), this immediately implies that
the parallel mobility always exceeds the perpendicular mobility, $\mu_{\parallel}>\mu_{\perp}$.
Furthermore, if both mobility coefficients are positive - as required
physically - their ratio necessarily satisfies $\mu_{\parallel}/\mu_{\perp}>1$.

\begin{figure}
\includegraphics[width=8.5cm]{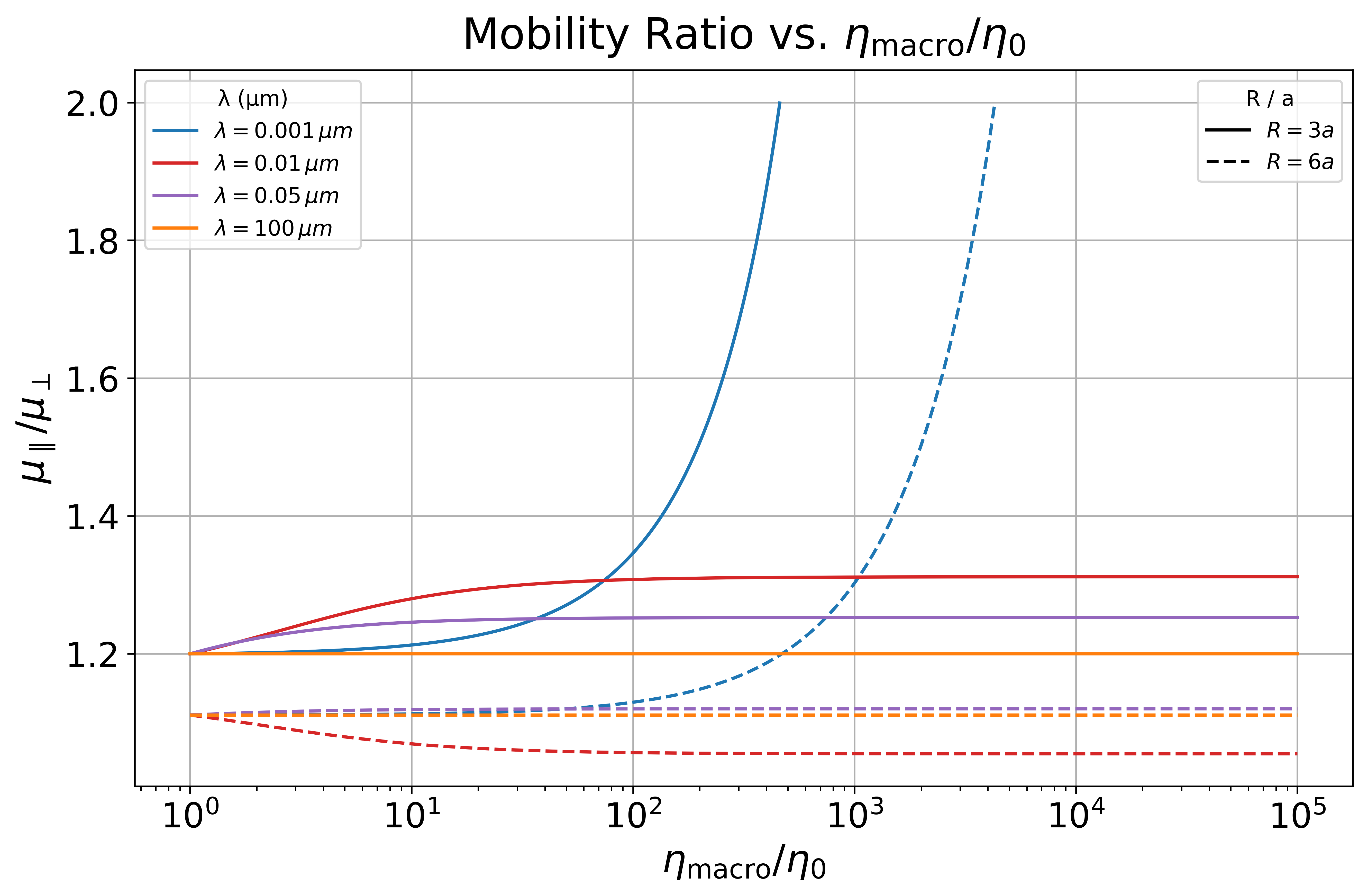}

\caption{Mobility ratio $\mu_{\parallel}/\mu_{\perp}$ for a dimer composed
of two identical spherical beads of radius $a=10^{-8}\,\mathrm{m}$,
separated by a center-to-center distance $R$. The mobility coefficients
are determined by the ``dominant contribution approximation'' defined
by Eq. (\ref{eq:dominant=000020contribution=000020app}) in the complex
fluid with the viscosity function from Eq. (\ref{eq:=000020rational=000020eta(k)})
characterized by macroscopic-to-solvent viscosity ratio $\eta_{\mathrm{macro}}/\eta_{0}$
and length $\lambda$ $\left[\mu m\right]$. The single-bead mobilities
required in the ``dominant contribution approximation'' were determined
from Eq. (\ref{eq:phi=000020within=000020Soft=000020Matter}).}\label{fig:mobility-raitos}
\end{figure}

The dumbbell mobility depends on four parameters $a$, $R$, $\eta_{\mathrm{macro}}/\eta_{0}$,
and $\lambda$. The first two characterize the dimer, while the latter
two characterize the viscosity function of the complex fluid. Fig.
\ref{fig:mobility-raitos} shows the mobility ratio curves for a dimer
composed of two identical spherical beads of radius $a=10^{-8}\,\mathrm{m}$,
separated by a center-to-center distance $R=3a$ and $R=6a$. In our
calculations, the macroscopic viscosity was varied over the range
$\eta_{\mathrm{macro}}/\eta_{0}\in[1,10^{5}]$. The length $\lambda$
spanned values from $10^{-3}\,\mathrm{\mu\mathrm{m}}$ to $10^{2}\,\mathrm{\mu\mathrm{m}}$,
corresponding to different complex fluids. For a dimer with $R=3a$
and $a=0.01\,\mu\mathrm{m}$, with $\lambda\in[0.001,100]\,\mu\mathrm{m}$
we observe an increase of the mobility ratio with $\eta_{\mathrm{macro}}/\eta_{0}$,
which sometimes reaches a large ratio $\mu_{\parallel}/\mu_{\perp}\gg1$.
In this case, the parallel motion of the dumbbell significantly exceeds
the perpendicular mobility, resembling reptation-like motion of a
polymer moving between immobile obstacles \citep{de1971reptation}.
Such reptation-like behavior can already be anticipated from the effective
Green function itself: as shown in Fig. \ref{fig:Velocity=000020fields}b,
the velocity field at distances of order $0.5\mu\mathrm{m}$ from
the point force exhibits regions of upward flow that effectively slow
transverse motion of the dumbbell. We note that reptation-like behavior
may also arise from structural features of complex fluids and direct
interactions, which, though important, are not the focus of the present
work. We performed the same analysis for a dimer with $R=6a$. While
for $\lambda=0.001\,\mu\mathrm{m}$ the mobility ratio increases,
for $\lambda=0.01\,\mu\mathrm{m}$ we observe a decrease of $\mu_{\parallel}/\mu_{\perp}$,
signaling nearly isotropic, sphere-like motion of the dimer.

We further determined the mobility of a dumbbell in aqueous polymer
solutions containing linear poly(ethylene glycol) chains of molecular
weight $M_{w}$ and concentration $c$ \citep{doi:10.1021/nl2008218}.
In these polymer solutions, phenomenological formulas for the probe
diffusivity as a function of probe radius, $\mu_{\text{single}}\left(a\right)$,
are available (see Eqs. (4--7) in Ref. \citep{doi:10.1021/nl2008218},
which we use with parameters $b=0.24$, $\beta=-0.75,$ $\alpha=0.62$;
notation as in Ref. \citep{doi:10.1021/nl2008218}). Applying these
in Eq. (\ref{eq:phi=000020within=000020Soft=000020Matter}) together
with Eqs. (\ref{eq:psi(R)=000020general}), (\ref{eq:mu=000020parallel})
and (\ref{eq:mu=000020perp}) yield the dumbbell’s parallel and perpendicular
diffusivities. We performed calculations for polymer solutions with
macroscopic viscosities $\eta_{\mathrm{macro}}/\eta_{0}\in[1,10^{5}]$,
molecular weights $M_{w}=\{325,\,3461,\,10944,\,15040,\,276862,\,2000000\}\,\mathrm{g/mol}$,
and the single-bead radius $a=0.01\,\mu\mathrm{m}$ and dumbbell size
$R=3a,6a$. These calculations in poly(ethylene glycol) lead to the
same conclusions as for the complex fluid described by the phenomenological
relation in Eq. (\ref{eq:=000020rational=000020eta(k)}): depending
on the parameters $c$, $M_{w}$, $a$ and $R$, a point force generates
a velocity field without (as in Fig. \ref{fig:Velocity=000020fields}a)
and with (as in Fig. \ref{fig:Velocity=000020fields}b) vortex-like
structures. Moreover, the diffusion of a dumbbell is either isotropic
($\mu_{\parallel}/\mu_{\perp}\approx1$) or resembles reptation-like
motion $(\mu_{\parallel}/\mu_{\perp}\gg1)$, similarly to the behavior
shown in Fig. \ref{fig:mobility-raitos}.

\section{Conclusions}

We have examined the diffusion of rod-like particles in complex fluids
using an approximation motivated by recent statistical-mechanical
insights into Smoluchowski dynamics. The key ingredients of the approach
are the mobility of a single spherical bead and the wave-vector-dependent
viscosity $\eta\left(k\right)$, which encodes dissipation across
length scales. Within this framework, the parallel and perpendicular
mobilities of a dimer follow from a simple expression involving the
effective Green function of the fluid.

This minimal description captures the full range of behaviors observed
for rod-like probes, from nearly isotropic, sphere-like motion to
strongly anisotropic, reptation-like dynamics. Because the approximation
depends only on $\mu_{\text{single}}$ and $\eta\left(k\right)$,
it can be applied broadly in experiments and simulations whenever
these quantities are known, without introducing additional phenomenological
parameters.

The range of conditions under which the hydrodynamic contribution
considered here dominates remains an open question. Clarifying these
limits will require further analysis.
\begin{acknowledgments}
W.S., H.J. and K.M. acknowledge support from the National Science
Centre, Poland, under Grant No. 2021/42/E/ST3/00180.
\end{acknowledgments}

\bibliographystyle{unsrt}

\end{document}